\documentclass[aps,pra,twocolumn,showpacs,superscriptaddress]{revtex4-1}

\usepackage{graphics}
\usepackage{graphicx}
\usepackage{dcolumn}
\usepackage{bm}
\usepackage{color}
\usepackage{amssymb,amsmath}

\begin{document}

\title{Guiding 2.94 $ \mu $m using low-loss microstructured antiresonant triangular-core fibers}

\author{Yang Chen}
\author{Mohammed F. Saleh}
\affiliation{Scottish Universities Physics Alliance (SUPA), School of Engineering and Physical Sciences, Heriot-Watt University, EH14 4AS Edinburgh, UK}
\author{Nicolas Y. Joly}
\affiliation{Max Planck Institute for the Science of Light, G\"{u}nther-Scharowsky str. 1, 91058 Erlangen, Germany}
\affiliation{Universit\"at Erlangen-N\"urnberg, Staudtstra\ss e 7/B2, 91058 Erlangen, Germany}
\author{Fabio Biancalana}
\affiliation{Scottish Universities Physics Alliance (SUPA), School of Engineering and Physical Sciences, Heriot-Watt University, EH14 4AS Edinburgh, UK}

\date{\today}

\begin{abstract}
We introduce a new simple design of hollow-core microstructured fiber targeted to guide mid-infrared light at a wavelength 2.94 $ \mu $m. The fiber has a triangular-core supported via silica-glass webs enclosed by a large hollow capillary tube. The fiber specific dimensions are  determined  based on the  guiding technique, which is based on the anti-resonant mechanism. For a triangular-core with side length $100$ $ \mu $m, the fiber has a minimum transmission loss  $0.08\pm0.005$ dB/m and dispersion 2.3 ps/km/nm at the operational wavelength 2.94 $ \mu $m.
\end{abstract}

\maketitle

\section{Introduction}
High-power laser sources such as Er:YAG laser at 2.94 $ \mu $m and CO$ _{2} $ laser at 10.6 $ \mu $m have influential medical and technological applications. The delivery of output intense pulses of these sources in silica-glass step-index fibers has always been a huge obstacle because of the low-damage threshold of the glass materials as well as the high attenuation of silica in the mid-infrared regime.

Hollow-core fibers (HCFs), with their unconventional guiding properties, seem to be the best solution for this problem  \cite{Ouzounov03,Luan04,Gerome07}. These fibers are engineered waveguides that use non-traditional methods for transferring light over relatively-long distances in a low-index core surrounded by a high-index structure. Photonic bandgap  (PBG) fibers are periodic microstructures that exploit the analogy with solid-state physics for confining light \cite{Knight96}. Light with  optical frequencies lying within a forbidden bandgap is trapped inside the  low-index-core due to the prohibited propagation inside the fiber-cladding. 1D and 2D PBG fibers have been demonstrated by wrapping a multilayered Bragg structure \cite{Yeh76,Johnson01,Abouraddy07} or stacking a number of capillary tubes in a suitably shaped preform \cite{Cregan99,Russell03,Knight03,Russell06}, respectively. By increasing the number of cladding layers, the fiber guiding loss is dramatically suppressed, albeit over restricted wavelength bands.  

Anti-resonant (AR) HCF fiber is another type of microstuctured fiber, where the guiding mechanism is based on the anti-resonant modes of Fabry P\'{e}rot cavities \cite{Duguay86,Litchinitser02}. These fibers have wide transmission windows, but with higher losses in comparison to the PBG fibers. Also, adding more layers to the cladding has small influence on decreasing the fiber-losses \cite{Ding14}. In recent years, a great deal of effort has been dedicated to suppress dramatically the losses of AR-HCF far beyond that of PBG fibers, by using multi-non-concentric capillary tubes \cite{Belardi14}. Contact points between cladding capillaries have to be avoided, since they would break the antiresonance condition and may lead to additional losses \cite{Kolyadin13}. Another technique for loss reduction is to add an extra outer cladding with an optimum displacement from the core \cite{Poletti11}. Particular negative curvature cores have been also shown to substantially lessen the fiber-losses \cite{Wang10,Pryamikov11,Belardi13}. In this case, careful design is required to prevent coupling between the core and cladding modes \cite{Belardi13}. Very recently, broadband single-mode operation have been demonstrated in these fibers by  resonant filtering of higher order modes \cite{Guenendi15}.

\begin{figure}\centering
\includegraphics[width=5cm]{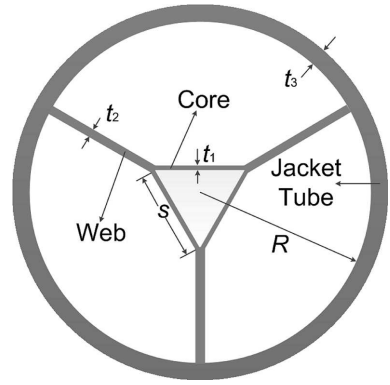}
\caption{(Color online). Sketch of our proposed hollow-core anti-resonant triangular-core fiber. $t_1$, $t_2$, and $t_3$ are  the cladding, web, and jacket tube thickness, respectively, $s$ is the core-side length, and $R$ is the jacket tube radius. \label{Fig1}}
\end{figure}

\section{Fiber design}
The delivery of the output light emitted by Er:YAG laser at 2.94 $ \mu $m was achieved by using tube-leaky fibers with losses 0.85 dB/m \cite{Kobayashi14}, or hollow-core negative-curvature fibers with losses 0.06 dB/m \cite{Urich13}. Also as recently predicted, transmission losses at this particular wavelength can reach minimal values using AR fibers with large numbers of nested tubes arranged in a certain order \cite{Habib15}. In this paper, we present an AR-HCF with a simple design for guiding mid-infra red light at 2.94 $ \mu $m with only $0.08\pm0.005$ dB/m losses. The fiber is made of three identical adjacent silica-glass capillary tubes encircled by a wider capillary with radius $ R  $ acting as a jacket tube with thickness $ t_{3} $. During the fabrication process the three touching capillaries create a perfect equilateral triangle ought to the action of surface tension in absence of additional external pressure, as shown in Fig. \ref{Fig1}. The triangle has a thickness  $ t_{1} $  determined by the dimension of the original capillary, and a length  $ s=R(1+2/\sqrt{3})^{-1}  $. The thickness $ t_{2} $ of the supporting webs is constrained by $ t_{2}=2 t_{1} $ due to fabrication requirements. It has been shown recently that a core with a polygon shape will have lower guiding losses as the number of the polygon sides decreases \cite{Ding14}. Hence, one can conjecture that our fiber would have lower losses in comparison to another one  with a hexagonal core, for instance. The fiber is anticipated to have great impact on the delivery of high-power laser pulses \cite{Ouzounov03,Luan04,Gerome07} as well as in gas-based nonlinear optical applications \cite{Russell14}.

\begin{figure}
\includegraphics[width=8.6cm]{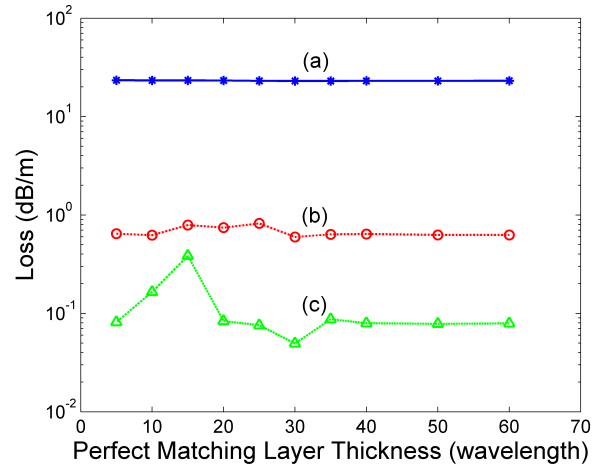}
\caption{(Color online). Loss-dependence  of hollow-core anti-resonant triangular-core fibers on the thickness of the perfect matching layers with $ t_{1} =  t_{3} = $ 2.19 $  \mu $m and $ \lambda = $ 2.94    $\mu $m. Blue stars, red circles, green triangles represent the fibers (a), (b), and (c).\label{Fig7}}
\end{figure}

The simulations in this paper are done using COMSOL software for three different fibers with core side lengths $s=$40, 70, 100  $ \mu $m, which we rename so forth as (a), (b), and (c), respectively. A suitable perfectly matched layer (PML) is used to estimate the attenuation of each fiber.  PMLs with enough thicknesses, several times the operating wavelength, have been used for the fibers (a), (b), and (c), respectively. Convergence of losses with PML thickness for the three fibers has been achieved as depicted in Fig. \ref{Fig7}.  For instance, the attenuation of fiber (c) is  $0.08\pm0.005$ dB/m at the designing wavelength 2.94 $ \mu $m. The error is deduced from the slight oscillations during numerical convergence, observed when increasing the PML layer to large values. Material absorption losses have not been included similar to recent studies \cite{Yu13}. A capillary tube that has the thickness of the triangle (c) and the radius of its inscribed circle has losses 26.86 dB/m, which demonstrates the effectiveness of our design.

\begin{figure}
\includegraphics[width=8.6cm]{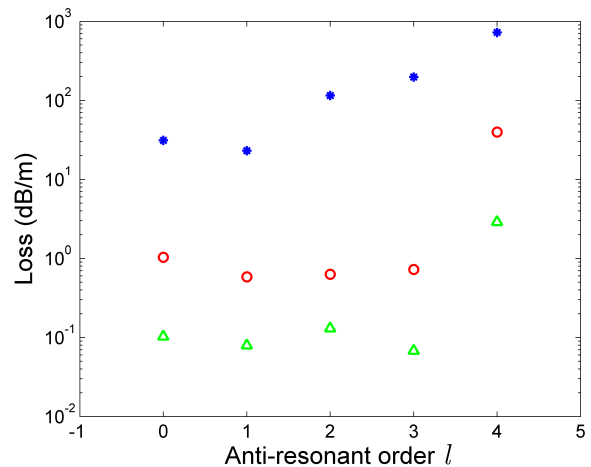}
\caption{(Color online). Losses of hollow-core anti-resonant triangular-core fibers with the core-side length $s=40$ $ \mu $m (blue stars), 70 $ \mu $m (red circles) and 100 $ \mu $m (green triangles) at the first five anti-resonant modes.\label{Fig2}}
\end{figure}

The anti-resonant wavelengths $ \lambda_{l} $ of an AR-HCF are given by \cite{Litchinitser02},
\begin{equation}
\lambda_{l}=\dfrac{4 t\sqrt{n^{2}_{2}-n_{1}^{2}}}{2l+1}, 
\label{Eq1}
\end{equation}
where  $ t $ is the cladding thickness, $ n_{1}$ and  $n_{2} $ are the core and cladding refractive indices, and $l\geq0$ is an integer that defines the order of the transmission window. Whereas the resonant wavelengths that leak outside the fiber-core during propagation are
\begin{equation}
\lambda_{m}=\dfrac{2 t\sqrt{n^{2}_{2}-n_{1}^{2}}}{m}, 
\label{Eq2}
\end{equation}
where  $ m $ is another positive integer, which represents the order of the loss peaks. Knowing the operating wavelength, different triangular-core thicknesses $ t_{1} $ can be determined using Eq. \ref{Eq1} by replacing $ t=t_{1} $, corresponding to the different values of $ l\geq0$. Based on the fact that the loss will grow with the rise of anti-resonant order, we have focused our attention on the first few values of $ l $. The guiding loss of the fundamental mode for fibers (a), (b), and (c) at the design wavelength 2.94 $ \mu $m for  $ l=0-4 $ are depicted in Fig. \ref{Fig2}.  For these values of $ l $, the corresponding thicknesses $ t_{1} $ are    0.73, 2.19, 3.64, 5.10, and 6.56  $ \mu $m, respectively. Based on the results shown in Fig. \ref{Fig2}, we have chosen the 1$ ^{\mathrm{st}}-$order anti-resonance with $ t_{1} =$ 2.19 $ \mu $m thickness, since it results in transmission losses $ < 0.1  $ dB/m for fiber (c). 

\begin{figure}
\includegraphics[width=8.6cm]{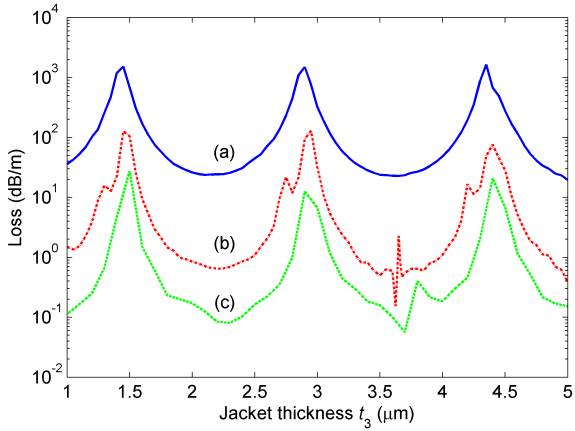}
\caption{(Color online). Loss-dependence of hollow-core anti-resonant triangular-core fibers on the jacket tube thickness $ t_{3} $, with $ t_{1} = $ 2.19    $\mu $m and $ \lambda = $ 2.94    $\mu $m. Straight blue, dashed-dotted red, and dashed green lines represent the fibers (a), (b), and (c) with side lengths $s=$ 40,  70, and  100 $ \mu $m.\label{Fig3}}
\end{figure}

The effect of the jacket tube thickness on the losses for the three fibers has been portrayed in Fig. \ref{Fig3} with  $ t_{1} = $ 2.19    $\mu $m. First, as the core-side increases, the core effective area also increases, and the loss will drop due to better mode accommodation inside the large core. In contrast to Ref. \cite{Yu13}, the losses in our fiber have a periodic oscillatory dependence as $ t_{3} $ varies. The loss minima occurs when the jacket tube thickness also satisfies the anti-resonant condition Eq. \ref{Eq1}. Moreover, the loss peaks are positioned at 1.46 , 2.91, and 4.37 $ \mu $m, which are exactly the thicknesses that fullfil the resonant condition Eq. \ref{Eq2} for $ m=1,2,3 $. Although light is trapped inside the triangular-core, satisfying this resonance condition in the jacket-tube ring allows the confined mode to escape from the core towards the jacket through the fiber webs.

\begin{figure}
\includegraphics[width=8.6cm]{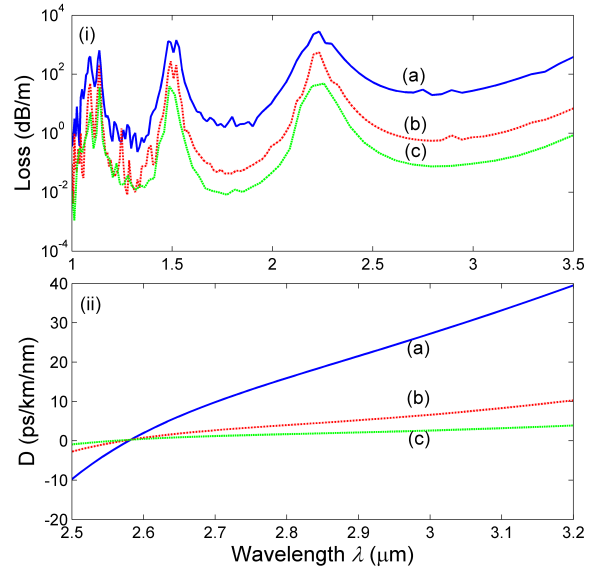}
\caption{(Color online). Wavelength-dependencies of (i) losses and (ii) dispersion of the triangular fibers (a), (b), and (c). The simulation parameters are  $ \lambda_{l}=$ 2.94 $   \mu $m, and $ t_{1} =  t_{3} = $ 2.19 $  \mu $m. \label{Fig4}}
\end{figure}

Our fiber performance measured by the spectral-dependence of its losses and dispersion is displayed in panels (i,ii) of Fig. \ref{Fig4} for the three fibers under consideration with $ t_{1} =  t_{3} = $ 2.19 $  \mu $m. The loss-dependence behaves similarly  to Fig. \ref{Fig3}, however, with slightly decreasing losses as moving towards shorter wavelengths. The reason is that the modal effective area shrinks when decreasing the wavelength, hence, the mode is better confined and its guiding-loss drops. The loss maxima and minima  take place when the wavelength matches  Eqs. (\ref{Eq1}) and (\ref{Eq2}), respectively. For the fiber (c), with side-length 100 $ \mu $m, the attenuation approaches notably only $0.08$ dB/m at the designing wavelength 2.94 $ \mu $m, very close to that value obtained using the negative curvature fiber \cite{Urich12}. As demonstrated in panel (ii), the fiber exhibits also flat low dispersion of about 2.3 ps/km/nm around this wavelength, especially for the case (c), allowing the fiber to be of great interest for nonlinear light-matter applications, especially when the core is filled with gases.

\begin{figure}
\includegraphics[width=8.6cm]{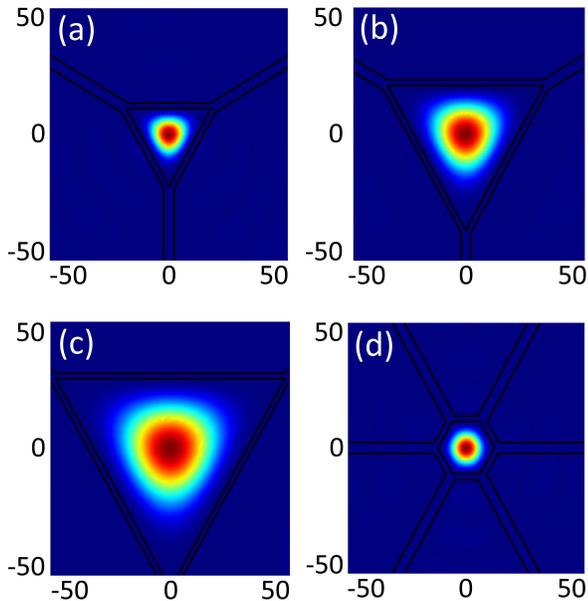}
\caption{(Color online). (i)--(iii) Fundamental-mode profiles of the triangular fibers (a), (b), and (c). The simulation parameters are  $ \lambda_{l}=$ 2.94 $   \mu $m, and $ t_{1} =  t_{3} = $ 2.19 $  \mu $m. (iv) Fundamental mode profile of a hexagonal fiber with . The values of the effective areas of each mode are enclosed in the panels.\label{Fig6}}
\end{figure}

The fundamental mode profiles of the triangular-core fibers are displayed in the first three panels of Fig. \ref{Fig6} using the aforementioned designed parameters. The effective mode areas of the fibers (a), (b) and (c) are 133, 422, and 935 $\mu$m$^2$, respectively. In panel (d), we have compared the triangular fiber (c) to a hexagonal fiber that has the criterion of having the same incircle  enclosed by the triangle-core as well as the same jacket tube radius $R$. We have found that the losses of the hexagonal fiber is 156 dB/m, which is much higher than the  triangular-core fiber, as we have anticipated earlier.

\begin{figure}
\includegraphics[width=8.6cm]{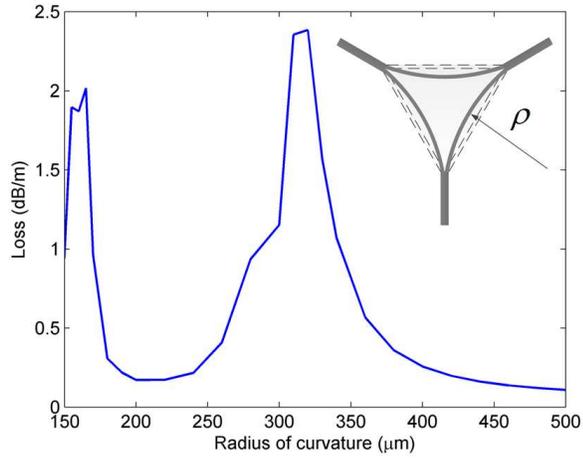}
\caption{(Color online). Loss-dependence of negatively-curved triangular-core fibers on the radius of curvature $\rho$  with fixed distance 100 $  \mu $m between its vertices. The simulation parameters are the same as in Fig. \ref{Fig4}. The inset shows a zoom of the geometry of the curved core. The jacket tube is still present, but is not shown in the figure. \label{Fig5}}
\end{figure}

We have also explored the possibility of implementing negative-curvature cores that can reduce the confinement loss more \cite{Wang10,Pryamikov11,Belardi13}. In designing this type of fiber, we maintain the locations of the three vertices as in the  straight case via choosing  a radius of curvature $\rho$ larger than the side of the triangle, $ \rho\geq s $. Using this method, the bent walls are prevented from touching each other, avoiding the formation of optical resonators that may lead to additional losses. In Fig. \ref{Fig5}, we have scanned the radius of curvature $ \rho $ from 150 to 500 $  \mu $m, and have calculated the losses. In this simulation, the determined values of the triangular-core and jacket thicknesses from the above studies have been used. Surprisingly, the losses of our fiber do not behave traditionally by varying the radius of curvature. For large values of $ \rho $, which in the limit $\rho\rightarrow\infty$ would correspond to the straight wall case,  the losses approach asymptotically the aforementioned values  0.08 dB/m.  With decreasing $ \rho $ or in other words increasing curvature, losses start to have strong peaks around $ \rho= $ 160 and 320 $  \mu $m due to large coupling overlap between the core and cladding modes \cite{Belardi13}. The minimum between these two peaks is approximately $\approx$ 0.2 dB/m, which indicates that negative curvature does not assist in suppressing the guiding losses.

\section{Conclusion}
In conclusion, we have proposed the design of a new simple  anti-resonant triangular-core fiber that can guide light at the mid-infrared frequency range. The fiber is made of relatively few identical capillary tubes with thickness 2.19 $ \mu $m. Our fiber has shown remarkably low-loss below 0.1 dB/m as well as low dispersion $\sim2.3$ ps/km/nm at  the operating wavelength 2.94 $   \mu $m. Also, we have shown that the thickness of the jacket tube has a significant role in confining the light inside the core. Moreover, we found that introducing negatively curved walls does not work towards loss-suppression. In fact, much higher losses have been obtained for certain values of the radius of curvature due to strong overlap between the modes of the core and the cladding. Finally, we believe that our design will induce other novel ideas and stimulate new research in the area of guiding far-infrared light.

M. Saleh would like to acknowledge the support of his research by Royal Society of Edinburgh and Scottish Government.

\bibliographystyle{apsrev4-1}	

%

\end{document}